\documentclass[aps,pra,twocolumn,amsmath, pxfonts,amssymb,nofootinbib,superscriptaddress]{revtex4}

\usepackage{graphicx}
\usepackage{epstopdf}

\newcommand{\bra}[1]{\langle#1|}
\newcommand{\perr}{p_{comp}}
\newcommand{\ploss}{p_{loss}}
\newcommand{\ket}[1]{|#1\rangle}
\newcommand{\eqn}[1]{Eq.~(\ref{#1})}
\newcommand{\fig}[1]{Fig.~\ref{#1}}
\newcommand{\logical}{computational }

\begin{document}
\title{Error Correction and Degeneracy in Surface Codes Suffering Loss}

\author{Thomas M. Stace} \email[]{stace@physics.uq.edu.au}
\affiliation{School of Mathematics and Physics, University of Queensland, Brisbane, QLD 4072, Australia}

\author{Sean D. Barrett\footnote{TMS and SDB contributed equally to this work.}} \email[]{seandbarrett@gmail.com}
\affiliation{Centre for Quantum Computer Technology, Macquarie University, Sydney, NSW 2109, Australia}

%
\date{\today}

\frenchspacing


\begin{abstract}
Many proposals for quantum information processing are subject to detectable loss  errors.  In this paper, we give a detailed account of recent results in which we showed that topological quantum memories can simultaneously tolerate both loss errors and computational errors, with a graceful tradeoff between the threshold for each.  We further discuss a number of subtleties that arise when implementing error correction on topological memories.  We particularly focus on the role played by degeneracy in the matching algorithms, and present a systematic study its effects on thresholds.  We also discuss some of the implications of degeneracy for estimating phase transition temperatures in the random bond Ising model.

\end{abstract}

\maketitle

\section{Introduction}

Quantum information is delicate. The state of a qubit can be corrupted by environmental noise, dissipation, or by imperfectly implemented logic operations on the qubit. Amongst the most remarkable achievements of quantum information science are the discoveries of quantum error correction (QEC) \cite{PhysRevA.52.R2493,PhysRevLett.77.793} and fault tolerant quantum computation (FTQC) \cite{steane1999eft}, which demonstrate that storage and manipulation of quantum information is possible, even in the presence of such imperfections. Typically, this protection is achieved by redundantly encoding the logical qubits in a larger number of physical qubits. 

A physical error may be classified as a \emph{\logical} error, in which the state of the qubits remains within the computational basis, or as a \emph{loss}, in which a physical qubit (e.g. a photon) is lost from the computer.
More generally, any detectable leakage process taking the qubit out of the computational basis can be treated as a loss error. Importantly, losses are presumed to be detectable, and in many qubit implementations, losses are a significant source of noise. 

Previous work by Dawson et al.\   \cite{dawson:020501} established thresholds against simultaneous loss and computational errors, finding that FTQC is possible provided  $\ploss \lesssim 0.3\%$ and  $\perr \lesssim 0.01\%$. Various studies, considering each error process in isolation, have derived much higher thresholds for each: Varnava et al.\ \cite{varnava:120501} constructed an encoding with a loss tolerance threshold of 50$\%$, whilst Knill \cite{knill-2004} and Raussendorf et al.\  \cite{1367-2630-9-6-199} have demonstrated computational error thresholds at the 1$\%$ level for FTQC.    In recent work considering the restricted problem of storing quantum information in \emph{surface codes} \cite{Bravyi98,kitaev2003ftq,dennis:4452}, we showed that these thresholds can be achieved in the presence of simultaneous losses and computational errors  \cite{stace:200501}.  

The properties of surface code stabilisers that enable this synthesis of the known loss and computational error thresholds for quantum memories are shared by the more elaborate schemes developed by Raussendorf et al. \cite{Raussendorf20062242,raussendorf:062313} that are capable of full FTQC.  We therefore fully expect that the methods we describe here will apply directly to FTQC schemes, including losses.

In parallel, a fruitful line of research has established the intriguing  connection between classical error correction and statistical mechanics models \cite{PhysRevLett.70.2968}.    More recently, similar connections between quantum error correction and phase transitions in many body systems have been made.   Notably, Dennis et al. \cite{dennis:4452} pointed out that optimally decoding an error syndrome on a surface code was equivalent to solving the classical 2-dimensional random-bond Ising model  (RBIM).  This observation implies that the error correction threshold coincides with the Nishimori point of the RBIM, i.e.\, that the error correction threshold is $\approx10.9\%$ \cite{wang2003cht,PhysRevB.65.054425}.  Further, numerical studies attempted to not only implement computationally efficient error correction algorithms, based on Edmonds' minimum weight matching algorithm, but to simultaneously make predictions about the phase diagram of the RBIM.  On the basis of large Monte Carlo simulations, Wang et al. \cite{wang2003cht} established very precise predictions about the location of the zero temperature phase transition in the RBIM, namely that $p_{c0}=10.31\%\pm0.01\%$.

The purpose of this paper is twofold.  Firstly, we expand on the discussion in  \cite{stace:200501} of a number of subtleties in the implementation of our loss tolerant protocol.  We provide a more detailed derivation of the edge weights that were introduced in  \cite{stace:200501}.  We then discuss the `phase-diagram' in the parameter space of losses and computational errors giving more detailed results of our numerical Monte Carlo simulations, as well as a new analytic result for slope of the phase boundary at small values of the loss rate.  We also give an exposition  of the interactions between error chains and percolated clusters of loss on finite lattices leading to a deviation from universal scaling as the loss rate approaches the percolation threshold.

Secondly, we discuss the role of degeneracy in matching algorithms for error correction and for the RBIM.  In the context of error correction, there are in fact very many matchings that attain the minimum weight, and so in this sense the code is highly degenerate.  This degeneracy was recently discussed in \cite{ducloscianci-2009}, and was implicit in the edge weighting introduced in \cite{stace:200501}. We derive expressions for the matching probability, accounting for path degeneracy.  We show that the degeneracy behaves analogously to an entropy term in a quasi-free energy.  On a square lattice, we provide a combinatorial expression for the shortest-path degeneracy of a given matching, and we exploit this to improve the error correction threshold for the surface code modestly, from $10.3\%$ \cite{wang2003cht} to above $10.6\%$.  

In the context of the RBIM, this phenomenon manifests itself in the very large degeneracy of the ground state manifold.   The degeneracy becomes important when computing numerical estimates of the locations of phase transitions in this, and related models.  In particular, we show that it is important to fairly sample from the ground state manifold in order to make accurate quantitative  predictions of phase transition thresholds.  We also give some evidence to suggest that existing implementations of matching algorithms do not do this, so may possibly suffer a systematic inaccuracy.

\section{Overview of the protocol}

For the purposes of analysis, the error model we consider is local and uncorrelated.  Each physical qubit is lost with probability $\ploss$. Losses are presumed to be detectable: a projector  onto the computational basis of a given qubit,  $\Pi_{i}=\ket{0}_i\bra{0}+\ket{1}_i\bra{1}$, is an observable indicating whether the state of the qubit has leaked out of the computational basis, without affecting the computational state of the qubit if it has not.  
 The remaining qubits are subject to independent bit-flip (X) and phase  (Z) errors, each with probability $\perr$. Both errors are handled in an analogous way in the surface code, so here we confine our attention to X errors, noting that the thresholds for Z errors will be identical. Aside from these errors, we assume other quantum operations (e.g. encoding operations, decoding operations, and syndrome measurements) can be implemented perfectly.

\begin{figure}
\begin{center}
\includegraphics[width=\columnwidth]{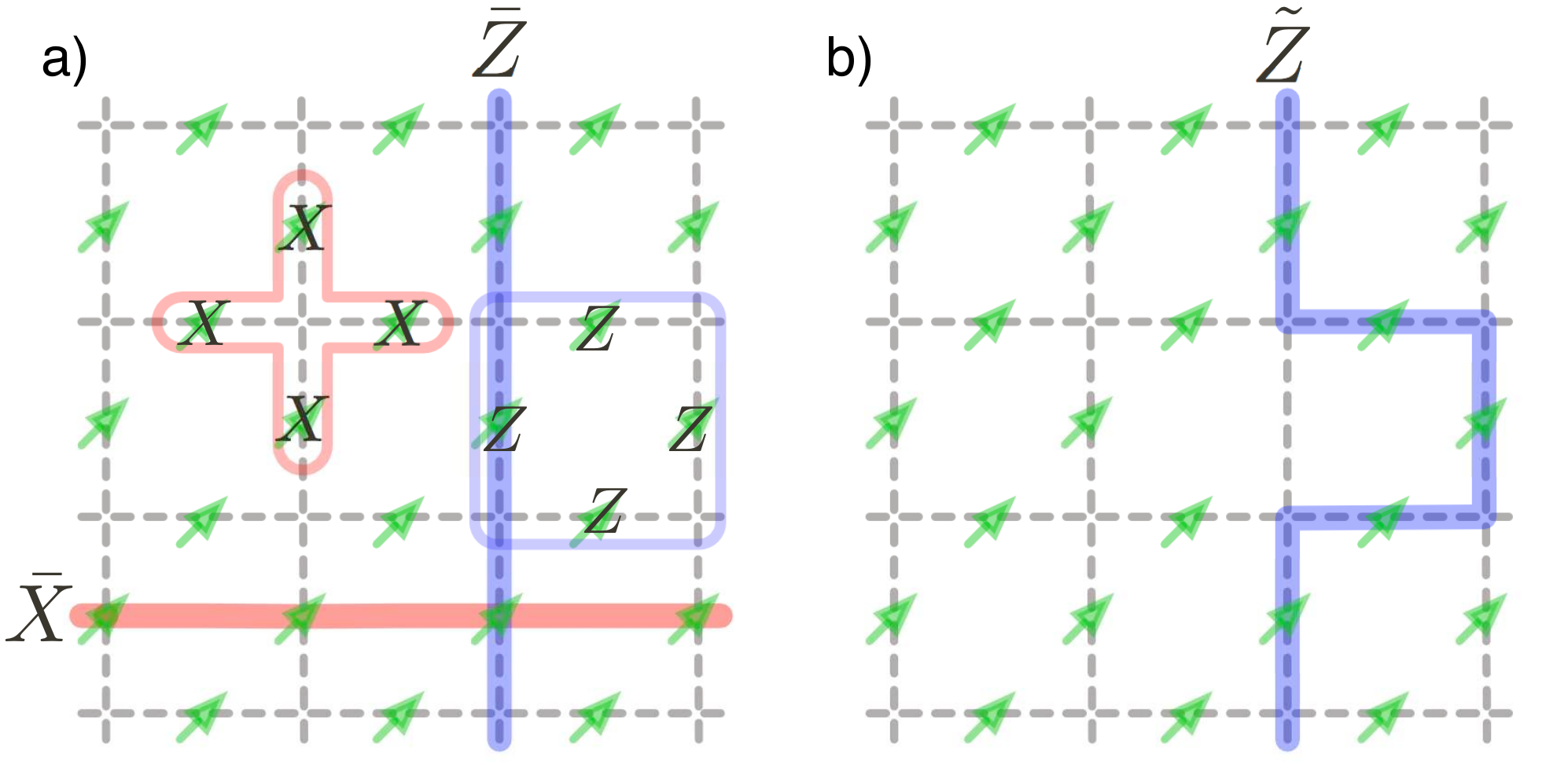}
 \caption{(a) Physical qubits (arrows) reside on the edges of a square lattice (dashed).  A plaquette operator is a product of four $Z$ operators acting on the qubits within the plaquette (blue), and a star operator is a product of four $X$ operators acting on the qubits within the star (red).  The logical $\bar Z$ operator is a product of $Z$ operators acting on qubits along the solid vertical (blue) line, and the logical $\bar X$ operator is a product of $X$ operators acting on qubits along the solid horizontal (red) line. (b) In the event of a qubit loss, an equivalent logical operator $\tilde Z$ can be routed around the loss.} \label{toriccode}
\end{center}
\end{figure}

Kitaev's surface codes are defined by a set of stabilisers \cite{PhysRevA.57.127}  acting on a set of physical qubits that reside on the edges of a square lattice.  The stabiliser group is generated by plaquette operators, which are products of $Z$ operators acting on qubits within a plaquette, $P_p =  \bigotimes_{i \in p} Z_{i}$, and star operators, which are products of $X$ operators acting on qubits within a star, $S_s = \bigotimes_{j \in s} X_{j}$, as depicted in \fig{toriccode} \cite{kitaev2003ftq}.  Star and plaquette operators all mutually commute, and the code space $\{|C\rangle\}$ is a simultaneous $+1$ eigenstate of all star and plaquette operators. If the $L\times L$ lattice has periodic boundary conditions (a torus), there are $2L^{2}$ physical qubits, $L^2-1$ independent plaquette operators and  $L^2-1$ independent star operators.  There are therefore two unspecified degrees of freedom, so the code is capable of encoding two logical qubits, $\bar q_i$  ($i \in \{1,2\}$).  A logical $\bar{Z}_i$ operator corresponds to a product of $Z$ operators along a homologically nontrivial cycle extending across the entire lattice, shown in \fig{toriccode}.  Likewise, a logical $\bar{X}_i$ operator corresponds to a cycle of $X$ operators extending across the entire lattice in the conjugate direction. 
$\bar{Z}_i$  and $\bar{X}_i$  commute with the stabilisers, but are not contained within the stabiliser group.



Note that we can obtain a new logical operator by multiplying the original one by a plaquette stabiliser,  $\tilde{Z}_i = P_p \bar{Z}_i$. The actions of $\bar{Z}_i$ and $\tilde{Z}_i$   on the code space are the same, since \mbox{$\tilde{Z}_i |C\rangle = P_p \bar{Z}_i |C\rangle =   \bar{Z}_i  P_p|C\rangle =   \bar{Z}_i |C\rangle$}. Thus there are many homologically equivalent cycles spanning the lattice with which to measure each logical qubit operator, as shown in \fig{toriccode}(b). This redundancy allows us to obtain the loss threshold for the case $\perr=0$: if only a few  physical qubits are lost, it is likely that each logical operator can still be measured by finding a homologically nontrival cycle which avoids all lost physical qubits. Thus the encoded quantum information can be reliably recovered. 

If  $\ploss$ is too high, there is likely to be a \emph{percolated} region of losses that spans the entire lattice, in which case there are no homologically nontrivial cycles with which to measure the logical operators. In the limit of large $L$, the boundary between recoverable and non-recoverable errors is a sharp transition corresponding to the \emph{bond percolation threshold} on the 2D square lattice \cite{Stauffer1985aa}: for $\ploss<0.5$ error recovery can almost surely be achieved, whereas for  $\ploss>0.5$ error recovery almost surely fails. Notably, this threshold saturates the fundamental bound on $\ploss$  imposed by the no-cloning theorem \cite{PhysRevLett.78.3217}.




The case  $\ploss=0$  and $\perr>0$ has been well studied \cite{kitaev2003ftq,dennis:4452,wang2003cht}. 
Briefly, physical bit-flip errors can lead to logical bit-flip ($\bar{X}_i$) errors but not logical phase  errors, and vice-versa. 
An \emph{error chain}, $E$, is a set of lattice edges (i.e.\ physical qubits) where a bit-flip error has occurred. $E$ changes the eigenvalues of the plaquette operators only at the \emph{boundary}, $\partial{E}$, of the chain to $-1$.  A connected chain of errors leads to a single pair of plaquettes with 
eigenvalue $-1$ at the endpoints of the chain. Measuring the plaquette operators therefore yields information about the endpoints of such connected chains. If the logical $\bar{Z}_i$ operator crosses the error chain an odd number of times, then the logical qubit suffers a $\bar{X}_i$ error. These errors may be corrected if the endpoints can be matched by a correction chain, $E'$, such that the closed chain $C=E+E'$ crosses the logical  $\bar{Z}_i$ chain an even number of times, i.e.\ is homologially trivial.  The error rate below which the correction chain $E'$ may be successfully constructed is closely related to the phase boundary of the random-bond Ising model (RBIM) \cite{wang2003cht,PhysRevE.64.046108}. If $\perr<p_{c0}=0.104$\footnote{Some studies report $p_{c0}=0.1031$ \cite{wang2003cht} whilst others report $p_{c0}=0.1049$ \cite{PhysRevE.64.046108}.}, then in the limit $L\rightarrow\infty$, the most probable chain, $C_{max}=E+E'_{max}$, is almost surely homologically trivial and recovery succeeds.  If $\perr>p_{c0}$, then in the limit $L\rightarrow\infty$, the chain is homologically trival only $25\%$ of the time.

\begin{figure}
\begin{center}
\includegraphics[width=0.9\columnwidth]{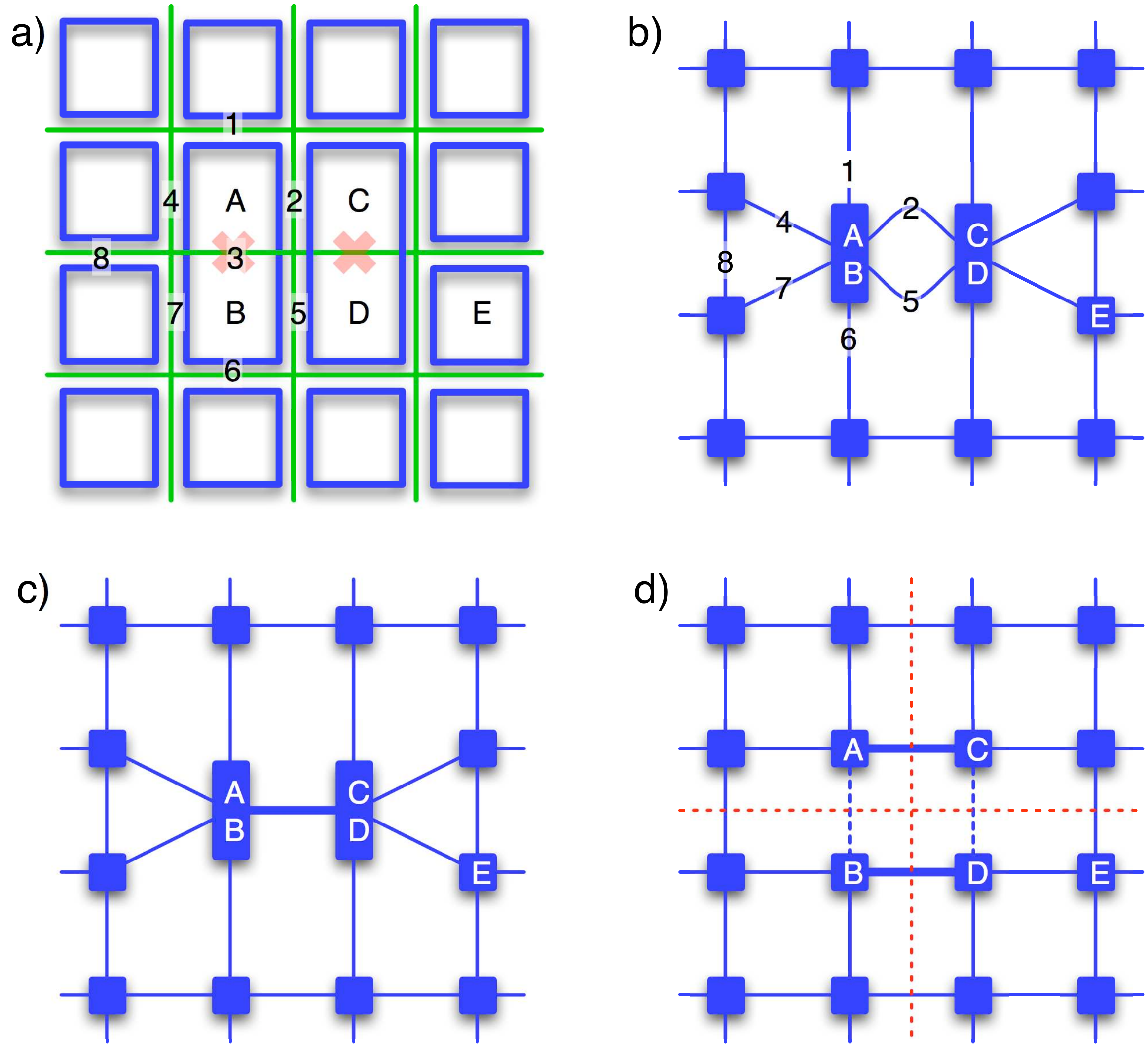}
\caption{(a) Lattice with two lost physical qubits.  The physical qubits are represented by the green lines between plaquettes.  Representative physical qubits are labelled $1...8$, and representative plaquettes are labelled $\mathsf{A, B, C, D, E}$.  (b) Connectivity of (super)plaquettes, where each edge represents a physical qubit. (c) Degraded lattice showing superedges (thick lines).  Black dots indicate the locations of Ising spins in the equivalent statistical mechanics model. (d) Restored lattice with zero weight edges (dashed) and edges between elements of superplaquettes with irregular weights (thick lines).  Also shown are the test lines for calculating the homology class of $C=E+E'$ on the restored lattice.} \label{multiedges}
\end{center}
\end{figure}

We can think of the above results as endpoints of a `boundary of correctability': $(\ploss,\perr)=(0.5,0)$ and $(0,0.104)$, respectively.   In the remainder of this paper, we demonstrate that toric codes (and planar codes, by extension) are robust against both loss and \logical errors, with a graceful tradeoff between the two.  We first describe how losses can be corrected by forming new stabiliser generators, which are aggregations of plaquettes or stars, called superplaquettes and superstars, respectively.  The superstar and superplaquette eigenvalues then reveal the error syndromes, and a  perfect matching algorithm  is used to find an error correction chain $E'$.
%
We illustrate the efficacy of the single round error correction protocol (i.e.\ ignoring fault-tolerance considerations) by calculating numerically the boundary of correctability in the $(\ploss, \perr)$ parameter space.



Consider the lattice shown in \fig{multiedges}(a) which is damaged by the loss of two physical qubits, marked by the red crosses.  The loss of qubit 3 affects two plaquette stabilisers: $P_\mathsf{A}=Z_1Z_2Z_3Z_4$ and $P_ \mathsf B=Z_3Z_5Z_6Z_7$, rendering them  unmeasurable.  
However, the \emph{superplaquette} $P_ \mathsf{AB}=P_ \mathsf AP_\mathsf B=Z_1Z_2Z_4Z_5Z_6Z_7$ is independent of the qubit at site 3, and so is a stabliser.  In the absence of errors, 
$P_ \mathsf{AB}$ has an eigenvlaue of $+1$. An error chain ending within the superplaquette $\mathsf {AB}$ changes the eigenvalue of $P_ \mathsf{AB}$ to $-1$.  
It follows that  the syndrome associated with a superplaquette is determined by the parity of the number of error chains that cross its boundary.  The fact that superplaquette operators yield syndrome information with which to construct an error correction chain, $E'$, is the basis for our loss-tolerant error-correction scheme.  

 In general, given any set of lost qubits, we can form a complete set of stabilisers on the damaged lattice in the following way: for each lost qubit $q$, which participates in neighbouring (super)plaquettes $P_q$ and $P'_q$, we form the superplaquette operator $P_qP'_q$, which is  independent of $Z_q$.  In the same way, we form superstar operators from products of star operators.  As discussed earlier, we can also form new logical $\tilde X_i$ and $\tilde Z_i$ operators by deforming the original logical operators to conform to the boundaries of newly formed superplaquettes.

We note that in \fig{multiedges}(a), there is a damaged plaquette operator $\bar Z_j=Z_3P_ \mathsf A=Z_1Z_2Z_4$ (or, equivalently $Z_3P_ \mathsf B=Z_5Z_6Z_7$) associated with the lost qubit 3,  which commutes with all the newly formed stabiliser generators on the damaged lattice, but whose eigenvalue, $\pm1$, is indeterminate.  Likewise, there is a damaged star operator $\bar X_j=X_4X_7X_8$ with indeterminate eigenvalue that also commutes with the new stabilisers on the damaged lattice.  Having indeterminate eigenvalues, these mutually-anticommuting, damaged operators form a two-dimensional degree of freedom in an unspecified state.  They therefore describe a completely mixed \emph{junk} qubit, $ j$,  which is a consequence of the entanglement between the lost qubit and the remaining qubits \cite{PhysRevA.71.022315}.  Since $\bar Z_{ j}$ and $\bar X_{ j}$ each commute with the new stabilisers, and with the deformed logical operators, the junk qubit is therefore in a product state with the logical qubits: $\ket{\psi}\bra{\psi}_{\bar q_1}\otimes\ket{\phi}\bra{\phi}_{\bar q_2}\otimes\mathbb{I}_{ j}/2$, and so the loss does not affect the logical qubits. 


When analysing the pattern of syndromes on the plaquettes and superplaquettes, we construct a new graph, depicted in \fig{multiedges}(b), in which a (super)plaquette is represented by a node, and (super)plaquettes share a bond on the new graph wherever they share a physical qubit in common.  Thus $P_ \mathsf{AB}$ and $P_ \mathsf{CD}$ share the qubits 2 and 5, and this is represented as two edges between the superplaquette nodes $\mathsf{AB}$ and $\mathsf{CD}$.

The error correction  syndrome $\partial E$ arising from an error chain on the graph in \fig{multiedges}(b) is determined by the (super)plaquettes that have an eigenvalue of $-1$.  To correct the errors, we follow the procedure described in previous work \cite{dennis:4452, wang2003cht} to find the most likely error chain giving rise to $\partial E$. 



\subsection{Super-edge Weights}

Calculating the probability of a given error chain is complicated by the presence of losses.  In the case where $\ploss=0$, the probability of an error on a qubit, $\ell=\{i,j\}$,  between two neighbouring plaquettes $i$ and $j$,  is uniform, $p_\ell=\perr$.    With losses, superplaquettes may share common physical qubits, as shown in \fig{multiedges}(b).  In this case,  the superplaquettes $\{\mathsf{AB},\mathsf{CD}\}$ have two qubits in common, 2 and 5, each of which might suffer an error. A non-trivial syndrome arises only if either qubit 2 or qubit 5 suffered an error, but not both.


By extension, for a pair of neighbouring superplaquettes, $\ell=\{P,P'\}$ sharing $n_{\ell}$ physical qubits, a non-trivial syndrome arises only if there are an odd number of errors on the $n_\ell$ qubits, which happens with probability
 \begin{eqnarray}
p_\ell&=&\sum_{m\textrm{ odd}}^{n_\ell} {n_\ell \choose m}\perr^m(1-\perr)^{n_\ell-m}\nonumber\\
&=&(1-(1-2\perr)^{n_\ell})/2.\label{edgeweight}
\end{eqnarray}

We therefore \emph{degrade} the graph shown in \fig{multiedges}(b), replacing multi-edges (i.e.\ several shared physical qubits) whose error probabilities are uniform,  with single \emph{superedges} whose error rates  depend on the number of physical qubits shared between neighbouring superplaquettes.  This degraded lattice is shown in \fig{multiedges}(c), in which there are no multi-edges, but the error probabilities are no longer constant.  Thus, the edge weights account for the degeneracy of possible error paths between neighbouring superplaquettes.  In subsequent sections, we will discuss the general issue of degeneracy on an arbitrary degraded lattice.

\subsection{Chain probabilities}

\begin{figure}[t]
\begin{center}
\includegraphics[width=0.7\columnwidth]{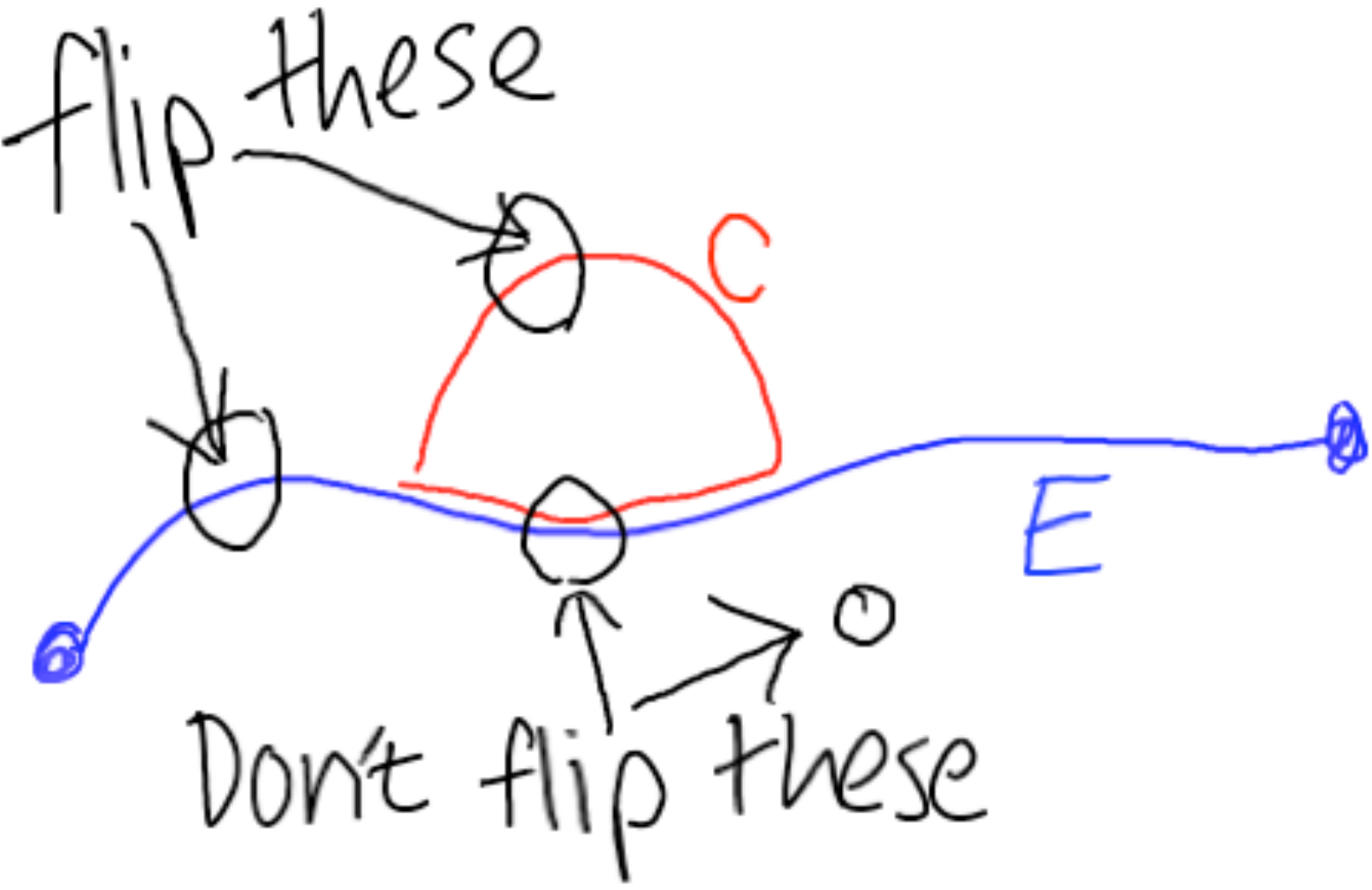}
\caption{Supposing the chain of errors $E$ to have been preciously laid out, the probability of then obtaining the path $C$ by further random flips, given by \eqn{probEprime}, is calculated is calculated with reference to this illustration.   The four black circles represent the regions over which the four products are taken in the first line of \eqn{probEprime}.} \label{transform}
\end{center}
\end{figure}

\begin{figure*}
\begin{center}
\includegraphics[width=18cm]{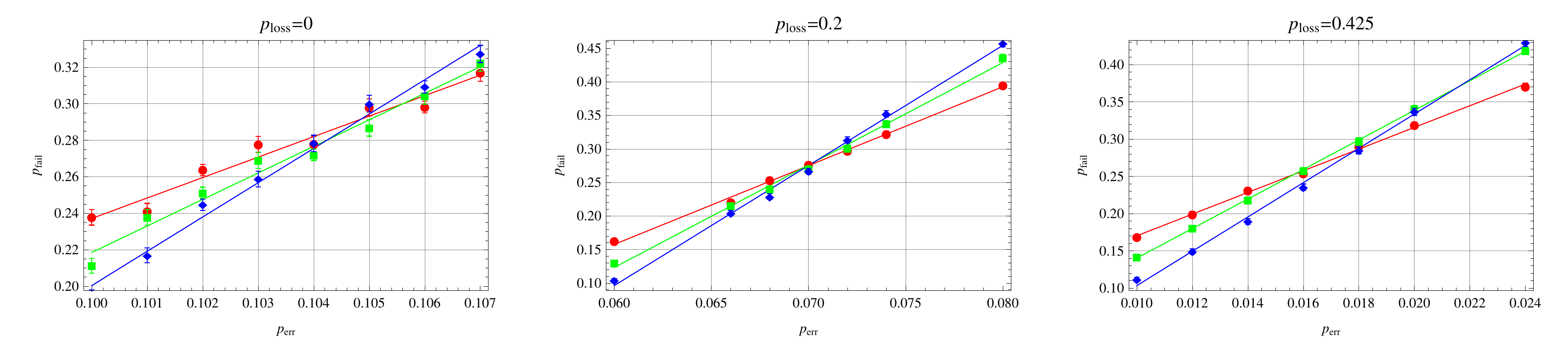}
\includegraphics[width=18cm]{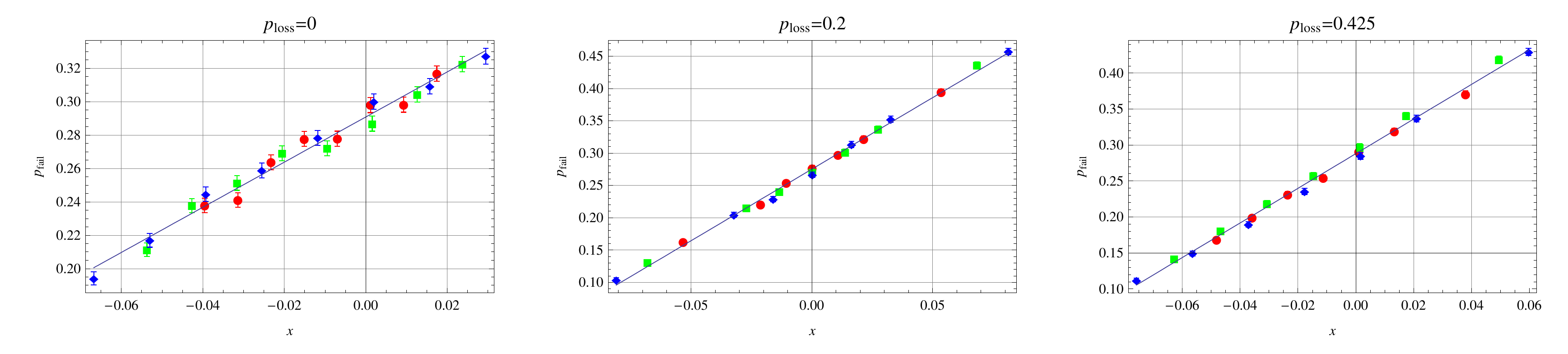}
\caption{(top) Statistical sampling of $p_{fail}$ vs.\ $p_{err}$. (bottom) Same data scaled according to $x=(p-p_{c0})L^{1/\nu_0}$, as in Harrington.  Then $p_{fail}(x)=a+b x$ is fitted to the scaled data.  Sample standard-error bars are shown.  Red circles are for $L=16$, green squares are for $L=24$ and blue diamonds are for $L=32$.  } \label{scaleddata}
\end{center}
\end{figure*}

Given a measured error syndrome indicating the boundary, $\partial E$, of errors on the degraded lattice, the probability of a given error chain $E'$ may be computed.  For completeness, and to define notation, we recapitulate the derivation of Wang et al.\ \cite{wang2003cht} (eq 15), generalised to include non-uniform edge weights.  Given an error chain $E$, we need a probability distribution for correction chains $E'$ that satisfy $\partial E'=\partial E$, that is we need $P(E'| E)$.  Since $C=E'+E$ (which happens to be a closed chain) it follows that $P(E'|E)=P(C|E)$.  The quantity $P(C|E)$ should be interpreted as ``Given the chain $E$ of flipped edges (i.e.\ qubits), what is the probability that independently flipping more edges results in \emph{only} the chain $C$".  The set of edges that would need to be randomly flipped to transform $E$ into $C$ is $\ell\in (C\cap\overline E)\cup(\overline C\cap E)$.  All other edges should remain unflipped, as depicted in  \fig{transform}.  The probability of this occurring is
\begin{eqnarray}
P(C|E)&=&\prod_{\ell\in C\cap E}(1-p_\ell)\prod_{\ell\in \overline C\cap \overline  E}(1-p_\ell)\prod_{\ell \in C\cap\overline E}p_\ell\prod_{\ell\in \overline C\cap E}p_\ell\nonumber\\
&=&\prod_{\forall}(1-p_\ell)\prod_{\ell\in(C\cap\overline E)\cup(\overline C\cap E)}\frac{p_\ell}{1-p_\ell}\nonumber\\
&=&\prod_{\forall \ell}(1-p_\ell)\left(\frac{p_\ell}{1-p_\ell}\right)^{n_\ell^C(1-n_\ell^E)+n_\ell^E(1-n_\ell^C)}\nonumber\\
&=&\prod_{\forall \ell}(1-p_\ell)\left(\frac{p_\ell}{1-p_\ell}\right)^{(1-u_\ell^Eu_\ell^C)/2}\nonumber\\
&=&\mathcal{N}\prod_{\forall\ell}\sqrt{\frac{1-p_\ell}{p_\ell}}^{\, u_\ell^{E}u_\ell^{C}}\nonumber\\
&=&\mathcal{N}\prod_{\forall\ell}e^{J_\ell u_\ell^{E'}}\label{probEprime},
\end{eqnarray}
where $\mathcal{N}=\prod_{\forall\ell}\sqrt{p_\ell(1-p_\ell)}$, $e^{J_\ell}=\sqrt{\frac {1-p_\ell}{p_\ell}}$ and for any chain $P$,
\begin{equation}
n_\ell^{P}=\left\{\begin{array}{cc}1 & \textrm{if }l\in P\\0 & \textrm{if }l\notin P\end{array}\right.\textrm{ and }u_\ell^{P}=\left\{\begin{array}{cc}-1 & \textrm{if }l\in P\\+1 & \textrm{if }l\notin P\end{array}\right.
\end{equation}
In the last line, we have used the identity  $u_\ell^{E'}=u_\ell^{E}u_\ell^{C}$, which follows from the fact that $E'=C+E$.   Note that $n_\ell^Q=(1-u_\ell^Q)/2$, 
and  that the definition of $J_\ell$ differs slightly from that in \cite{wang2003cht}.

On this degraded lattice with non-uniform edge weights, we may therefore assign a probability for any hypothetical chain $E'=E+C$.  This probability, which is conditioned on the syndrome $\partial E$, is  \cite{dennis:4452,wang2003cht}
\begin{equation}
P(E'|\partial E)=\mathcal{N}\prod_{\forall\ell}e^{J_\ell u_\ell^{E'}}.
\end{equation}

 Following \cite{wang2003cht}, $u_\ell^C$ satisfies the closed chain constraint $\prod_{\ell\in s}u_\ell^C=1$ for every (super)star $s$, which  can be `solved' by introducing the Ising spins, $\sigma_i$ and $\sigma_j$ on the dual lattice such that $u_{i,j}^C=\sigma_i\sigma_j$.  Then 
\begin{equation}
P(E'|\partial E)=\mathcal{N}\prod_{\langle   i j\rangle}e^{J_{ij} u_{ij}^{E}\sigma_i\sigma_j},\label{probE}
\end{equation}
and this probability is generated by the partition function for the RBIM
\begin{equation}
Z=\sum_{\sigma}\prod_{\langle   i j\rangle}e^{\beta J_{ij} u_{ij}^{E}\sigma_i\sigma_j},
\end{equation}
at $\beta=1$. 
There are thus two additional sources of randomness in this model compared to the standard RBIM: (a) the lattice connectivity is irregular, and (b) the magnitude of the Ising couplings, $J_\ell$ are irregular.  However, the probability distribution from which $u_\ell^E$ is selected and the magnitude of the coupling $J_\ell$ are both functions of the probability $p_\ell$.  Whilst this varies throughout the lattice,  each edge satisfies the a local Nishimori condition.

A ground state of this irregular RBIM ($\beta\rightarrow\infty$) can be found by maximIsing $P(E'|\partial E)$ over possible paths $E'$.  The path $E'$ that maximises $P(E'|\partial E)$ is the same as that which maximises $\ln P(E'|\partial E)$:
\begin{eqnarray}
\max_{E'}\ln P(E'|\partial E)&=&\max_{E'}\ln\prod_{\forall\ell}e^{J_\ell u_\ell^{E'}}\nonumber\\&=&\max_{E'}\sum_{\forall\ell}J_\ell u_\ell^{E'}\nonumber\\
&=&\max_{E'}\left(\sum_{\forall\ell}J_\ell-\sum_{\ell\in E'}2J_\ell\right)\nonumber\\
&=&\sum_{\forall\ell}J_\ell-\min_{E'}\sum_{\ell\in E'}2J_\ell\nonumber\\
&=&\sum_{\forall\ell}J_\ell-\min_{E'}\sum_{\ell\in E'}\ln\left(\frac{1-p_\ell}{p_\ell}\right).\nonumber
\end{eqnarray}
For the regular lattice, $p_\ell=p$, and so this amounts to finding the shortest path.  For a lattice with irregular weights (as we have), we should weight each edge by $\ln(({1-p_\ell})/{p_\ell})$, and find the path with the minimum additive weight.

The most likely error chain is given by $\max_{E'}P(E'|\partial E)$.  This  is equivalent to
\begin{equation}
\min_{E'}\sum_{\ell\in E'}\ln\left(\frac{1-p_\ell}{p_\ell}\right).
\end{equation}
This minimisation may be accomplished using Edmonds' minimum-weight, perfect-matching algorithm \cite{WilliamCook01011999}.  
For the case $\ploss=0$, this simply minimises the total metropolis length of the matching path, and is the same procedure implemented in previous studies \cite{dennis:4452, wang2003cht}.  In the case where $\ploss>0$, the edges do not have uniform weight, since $p_\ell$ depends on the number of physical qubits, $n_\ell$, shared between adjacent superplaquettes.

\subsection{Correctable Phase  Diagram}

In order to make quantitative estimates of the  tolerable error rates in the presence of loss, we take adopt two approaches.  Firstly we perform Monte Carlo simulations to provide statistical estimates of the boundary for all values of $p_{loss}$. Secondly, we provide a semi-analytic estimate for the threshold value of $p_{comp}$ as a function of $p_{loss}$.  These two approaches are in good agreement. 

\subsubsection{Monte Carlo Simulations}

For the purposes of simulation,  it is easier to determine homology classes on a square lattice, rather than the degraded lattice, exemplified in  \fig{multiedges}(c).  We therefore restore the square lattice by dividing superplaquettes into their constituent plaquettes in the following way:  (1) an edge between two plaquettes  within a single superplaquette is assigned a weight of zero, (2) an edge between plaquettes in two neighbouring superplaquettes is given the weight of the superedge in the degraded lattice, as illustrated in \fig{multiedges}(d).  These transformations do not change the minimum weighted-distance between any pair of syndromes, and so a minimum-weight perfect matching on the restored lattice is also a minimum-weight perfect matching on the degraded lattice.  Determining the homology class is then accomplished by counting crossings of  vertical and horizontal test lines in the dual lattice.

In order to test the efficacy of our loss-tolerant error correction scheme, we generate random losses on a periodic lattice with rate $\ploss$.   On the remaining edges we generate a set of errors, $E$, with rate $\perr$. Depending on the distribution of losses, we assign edge weights according to \eqn{edgeweight} to edges in the restored square lattice.  The syndrome, $\partial E$, is calculated, and applying Edmonds'  algorithm to $\partial E$  yields the maximum-likelihood error correction chain, $E'$.  The homology class of the chain $E+E'$ then determines whether error correction was successful.  


\begin{figure}
\begin{center}
\includegraphics[width=\columnwidth]{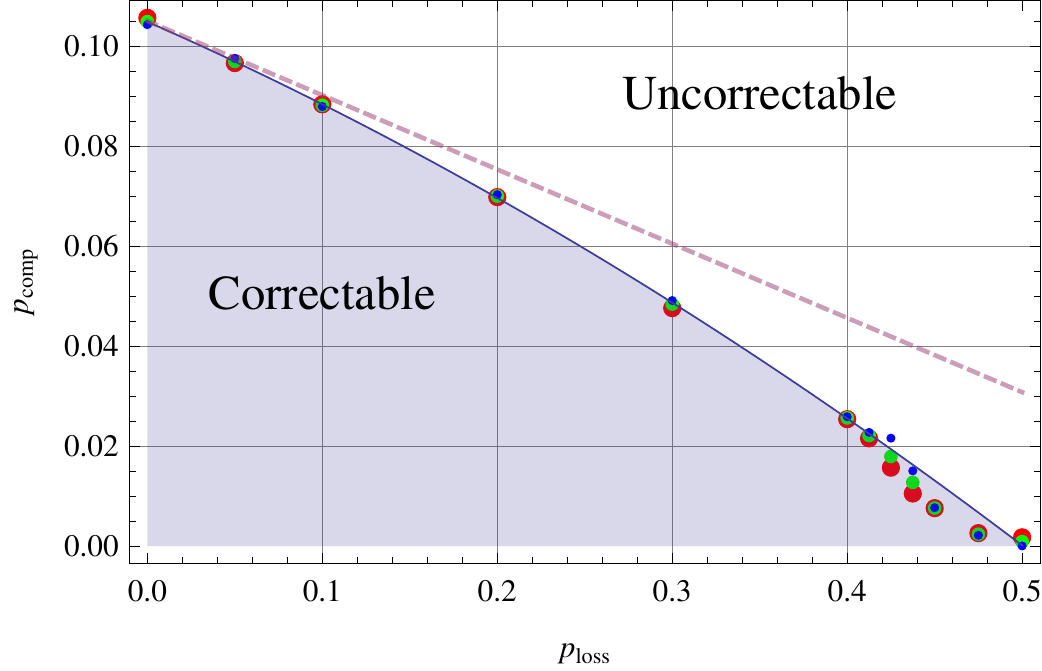}
\caption{ Correctability phase diagram.  The shaded region is correctable in the limit $L\rightarrow\infty$.  
Red points are the crossing point of $p_{fail}$ calculated on $L=16$ and $L=24$ lattices, green are the crossing of $p_{fail}$ (versus $\perr$) on $L=16$ and $L=32$ lattices, and blue are the crossing point of $p_{fail}$ on $L=24$ and $L=32$ lattices.  
The curve is a quadratic fit to the points in the interval $0\leq\ploss\leq0.4$ for which finite size effects are negligible. It extrapolates through the point $(\ploss,p_{comp})=(0.5,0)$.  The dashed line is the linearised phase boundary for $p_{loss}$ much less than the percolation threshold, given in \eqn{eqn:PBapprox}.} \label{PhaseDiagram}
\end{center}
\end{figure}

For a given value of $\ploss$ we simulate the protocol for different values of $\perr$ on different sized lattices, $L=16, 24$ and 32. 
For each value of $\perr$ and $L$, the failure rate, $p_{fail}$, is calculated by averaging over $10^4$ trials. 
Following \cite{wang2003cht}, we seek a threshold, $p_{comp}^{thr}$ (depending on $\ploss$), such that if $\perr<p_{comp}^{thr}$ then $p_{fail}\rightarrow0$ as  $L\rightarrow\infty$. Conversely, if  $\perr>p_{comp}^{thr}$ then $p_{fail}\rightarrow3/4$ as  $L\rightarrow\infty$.

 \begin{figure*}
\begin{center}
\includegraphics[width=0.6\columnwidth]{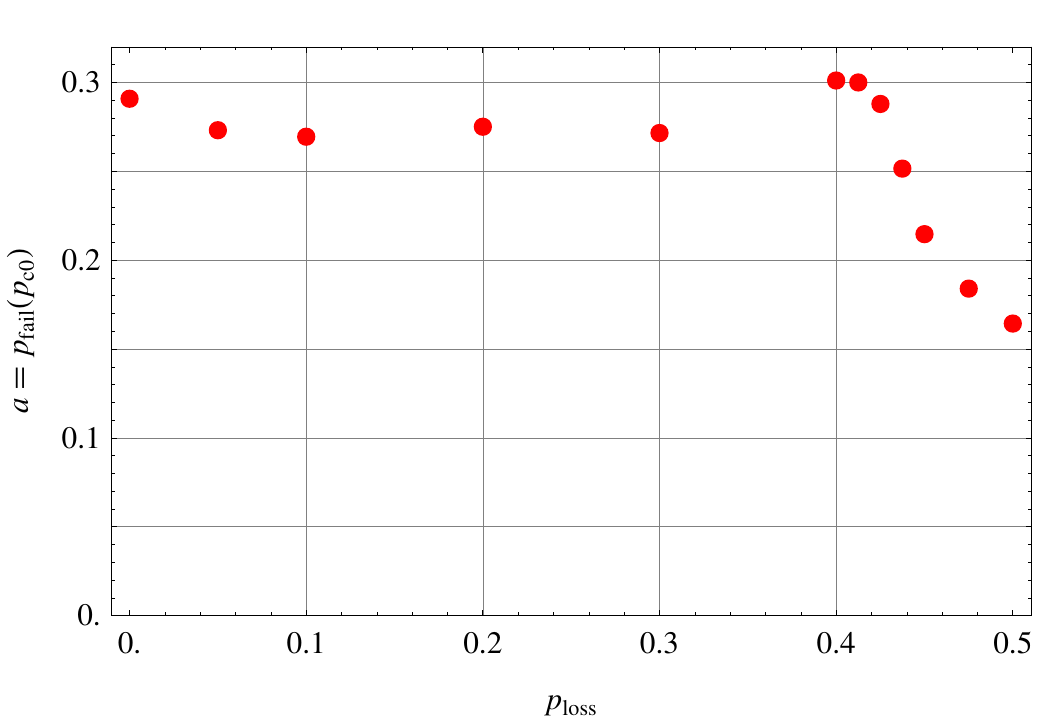}
\includegraphics[width=0.6\columnwidth]{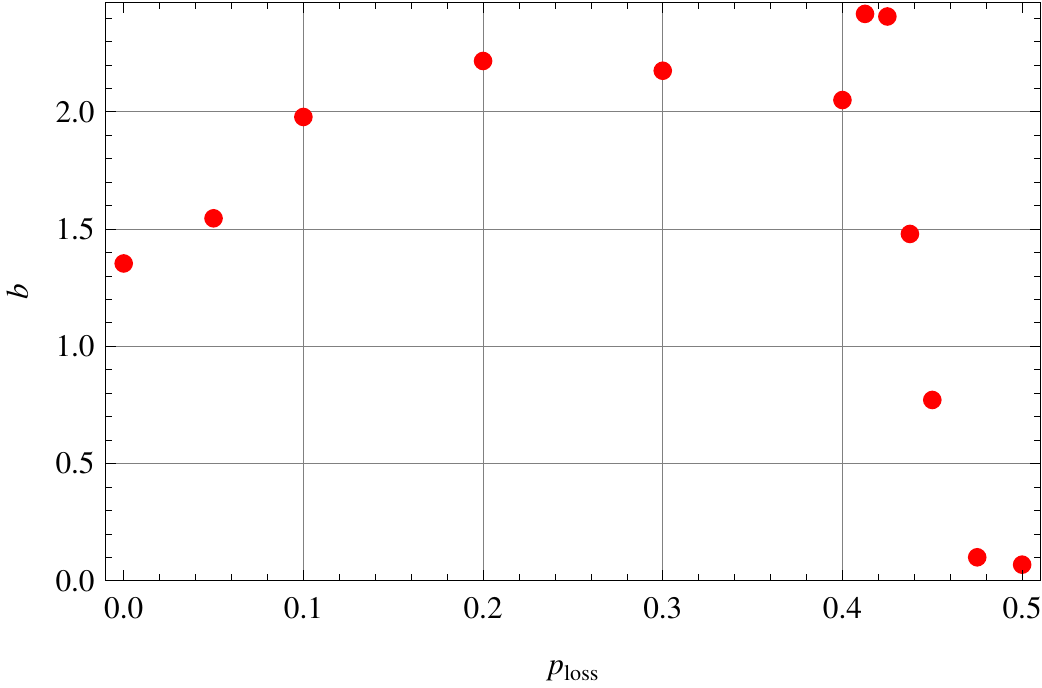}
\includegraphics[width=0.6\columnwidth]{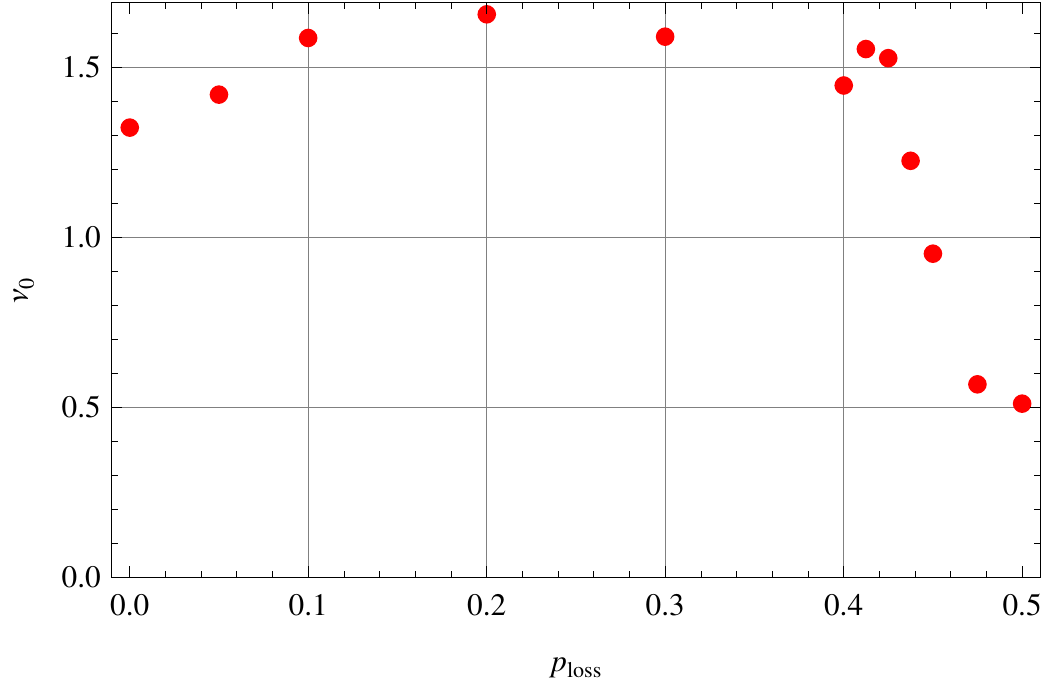}
\caption{Fit parameters, $a,b$ and $\nu_0$, for data collapse in \fig{scaleddata}(bottom).} \label{fitparameters}
\end{center}
\end{figure*}

Results of simulations for different values of $p_{loss}$ are shown in \fig{scaleddata}.  For each value of $p_{loss}$, the error rate is scaled according to the same prescription as given by Wang et al.:  $x=(p-p_{comp}^{thr})L^{1/\nu_0}$.  The scaled results generally fall on the same linear relationship, $p_{fail}=a+b x$, also shown. The parameters $p_{comp}^{thr}, \nu_0, a$ and $b$ with $p_{loss}$ and are determined by least squares fitting.  If there is universal scaling behaviour, then $p_{comp}^{thr}$ is independent of $L$, however at large values of $p_{loss}$, universal scaling breaks down, and the point at which the $p_{fail}$ curves cross depends on $L$.

\fig{PhaseDiagram} is the central result in this paper, and shows $p_{c0}$ as a function of $\ploss$.  Since we simulate three different values of $L$, there are in principle three different candidates for $p_{comp}^{thr}$, indicated by the different coloured/sized points.  For $\ploss\leq0.4$, these all coincide, so $p_{comp}^{thr}$ is well defined.  The failure rate obeys the universal scaling law $p_{fail}=f[(\perr-p_{comp}^{thr})L^{1/\nu_0}]$ 
 with scaling exponent, $\nu_0$, in the range 1.4 to 1.5, consistent with  the RBIM universality class \cite{wang2003cht}.    

Fitting a quadratic through the points in the interval $0\leq\ploss\leq0.4$ yields a curve that extrapolates through the point $(0.5,0)$.  This curve represents the boundary of correctability: if $(\ploss,\perr)$ is in the shaded region then error correction almost surely works, as $L\rightarrow\infty$.  Importantly, this boundary  passes through the known bounds at $\ploss=0$ and $0.5$, demonstrating that the protocol is very robust against loss.

\subsubsection{Linearised approximation for small $p_{loss}$}

If $p_{loss}$ is much less than the percolation threshold, so that losses are very sparsely distributed, then to a very good approximation, the only superplaquettes that arise are those consisting of two neighbouring plaquettes, and contain 6 physical qubits.  The fact that these superplaquettes contain more physical qubits than the original plaquettes means that they have a higher chance of suffering a syndrome error.  Therefore, as $p_{loss}$ increases, the the \emph{effective} rate of syndrome errors (averaged over plaquettes and superplaquettes) also increases.  We use this observation, that for a fixed computational error rate, a lossy lattice has a higher per-stabiliser error rate than a lossless lattice, to calculate the dependence of the error correction threshold on the loss rate, for small $p_{loss}$.

For small loss rates the probability that a given stabiliser has a syndrome error is, to a good approximation:
\begin{eqnarray}
P(\textrm{stab. error})&=&P(\textrm{no loss}) P(\textrm{plaq. error})\nonumber\\
&&{}+P(\textrm{1 loss}) P(\textrm{superplaq. error})\label{eqn:smalllosserror}
\end{eqnarray}
where  
\begin{eqnarray}
P(\textrm{no loss})&=&1-P(\textrm{1 loss})=(1-p_{loss})^4,\nonumber\\
P(\textrm{plaq. error})&=&(1-(1-2\perr))^4/2,\nonumber\\
P(\textrm{superplaq. error})&=&(1-(1-2\perr))^6/2\nonumber
\end{eqnarray}
are the probabilities that an odd number of physical qubits suffer an error on a plaquette (consisting of 4 qubits) and on a superplaquette (consisting of 6 qubits), respectively\footnote{The latter two are specific cases of \eqn{edgeweight}.}.  


The effective single qubit error rate, $p_{eff}$, on a \emph{lossless} lattice that duplicates the per-stabiliser error rate of the lossy lattice, \eqn{eqn:smalllosserror},  therefore satisfies
\begin{equation}
(1-(1-2p_{eff}))^4/2=P(\textrm{stab. error}).\label{eqn:implicit}
\end{equation}
From \cite{wang2003cht}, the threshold for this effective error rate on the lossless lattice is $p_{c0}=0.103$.  Thus, setting $p_{eff}=p_{c0}$ in the LHS of \eqn{eqn:implicit} yields an implicit equation relating the threshold error correction rate,  ${\perr^{thr}}$, to $p_{loss}$.  Treating $p_{loss}$ as the independent parameter determining the error threshold, so $\perr^{thr}=\perr^{thr}(p_{loss})$, we compute the slope 
$\alpha= {\perr^{thr}}'(0)$  by differentiating both sides of \eqn{eqn:implicit} with respect to $p_{loss}$.  The result is
\begin{eqnarray}
\alpha&=&-2\perr^{thr}(0)(1-\perr^{thr}(0))(1-2\perr^{thr}(0)),\nonumber\\
&=&-2p_{c0}(1-p_{c0})(1-2p_{c0}),\nonumber\\
&\approx&-0.148.\nonumber
\end{eqnarray}
It follows that the linear approximation to the phase boundary for small $p_{loss}$ is 
\begin{equation}
\perr^{thr}\approx p_{c0}+\alpha\, p_{loss}.\label{eqn:PBapprox}
\end{equation}
This relationship is shown as the dashed line in \fig{PhaseDiagram}.  

From the quadratic fit to the Monte Carlo results in \fig{PhaseDiagram}, $\alpha$ is found to be $-0.154\pm0.0033 \,(1\sigma)$.  This is in good agreement with the semi-analytic estimate given above.

\begin{figure*}
\begin{center}
\includegraphics[width=\columnwidth]{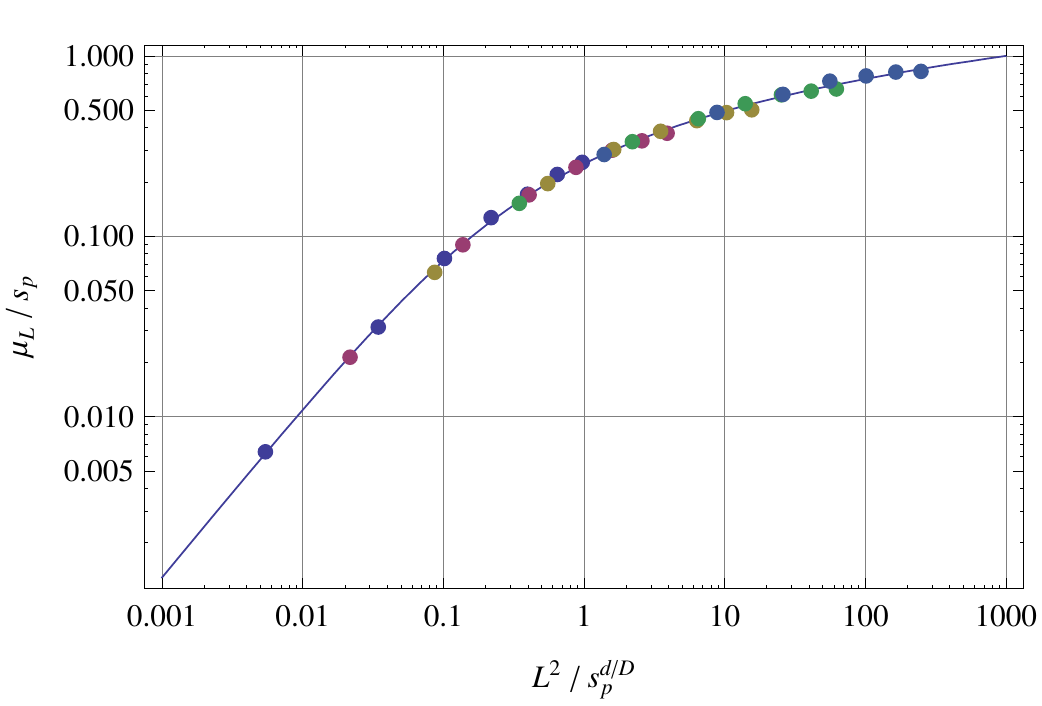}
\includegraphics[width=\columnwidth]{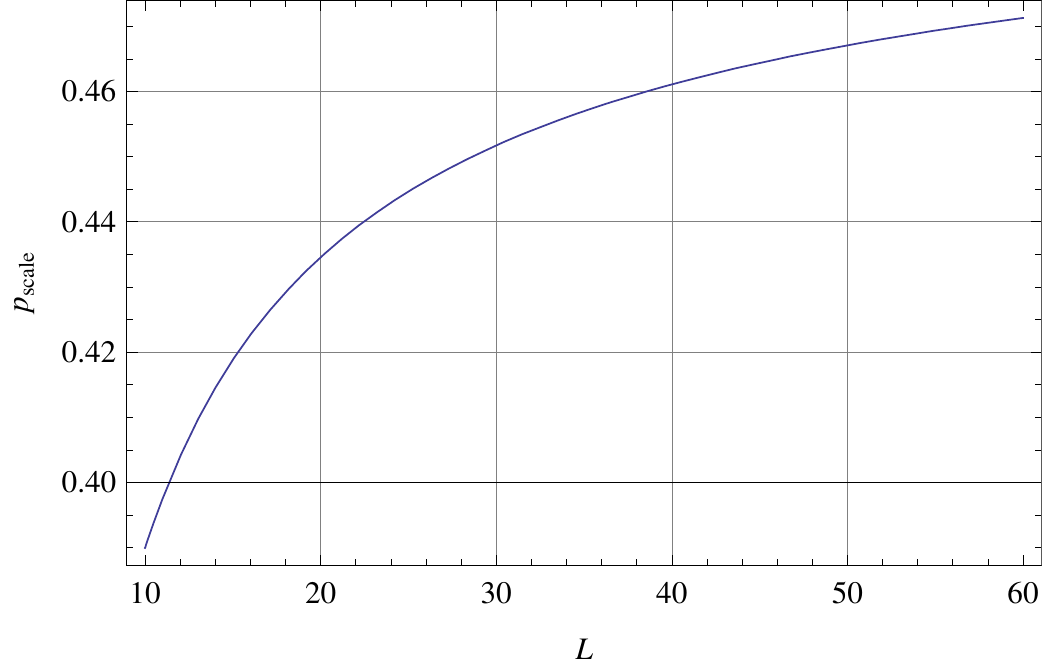}
\caption{(left) Scaling law for the size of the largest superplaquette on an $L\times L$ lattice.  $\mu_{L}$ is the number of sites in the largest superplaquette, $s_{p}$ is the number of sites within a `percolated-correlation length', $s_{p}=|p-p_{perc}|^{\nu D}$, where $p_{perc}=1/2$ is the percolation threshold for the lattice, $\nu=4/3$, is the correlation length scaling exponent, $D=91/48$  is the fractal dimension of the largest percolated cluster and $d=2$ is the dimensions of the lattice.  Points are statistical samples on different lattice sizes ranging from $L=2$ to 64, and the solid curve is a fitted scaling law, given by $\Phi(x)=0.116\log(7.74\,x^{D/d}+1)$ (adapted from \cite{PhysRevE.62.1660}). (right) Solving ${\mu_{L}(p_{scale})}=L^{2}/2$ for $p_{scale}$ as a function of $L$.  When $p_{loss}=p_{scale}$, the largest super-plaquette occupies half of the lattice.} \label{largestSP}
\end{center}
\end{figure*}

\subsection{Superplaquette Percolation and Finite Size Effects}

For $\ploss\geq0.425$, the universal scaling law breaks down, and the points in \fig{PhaseDiagram} lie below the quadratic extrapolation (though also reaching the point $(0.5,0)$).  This is attributed to the fact that for $\ploss\geq0.425$, the largest superplaquette on an $L\leq32\times32$ lattice  occupies approximately half of the lattice sites \cite{PhysRevE.62.1660}, so finite size effects dominate.  This can be seen in \fig{fitparameters}, in which the fit parameters $a,b$ and $\nu_0$ change precipitously for large values of $p_{loss}$ 

As $p_{loss}$ approaches $0.5$ (for fixed $L$), the size of the largest superplaquette becomes comparable to the size of the lattice, so it is reasonable to expect that the scaling behaviour at small $p_{loss}$ breaks down.  The size of the largest superplaquette, $\mu_{L}(p_{loss})$, on an $L\times L$ lattice is closely related to the size of the largest percolated cluster, discussed by Bazant \cite{PhysRevE.62.1660}.  We note that percolated cluster of losses may contain `islands' of qubits that were not lost.  For our purposes, these isolated islands within a cluster of losses neither assist in constructing deformed logical qubit operators, nor in creating superplaquettes.  Therefore, in analysing the size of percolated clusters, we treat these islands as if they are lost.  

\fig{largestSP}(left) shows $\mu_{L}$ versus $L^2$, with $\mu_{L}$ and $L^{2}$ are scaled by $s_{p}=|p-p_{perc}|^{\nu D}$ and $s_{p}^{d/D}$ respectively (where $\nu=4/3$ is the percolation length scaling exponent \cite{grimmett1999percolation,Stauffer1985aa}, $D=91/48$ is the fractal dimension of the largest percolated cluster \cite{PhysRevE.62.1660} and $d=2$ is the dimensions of the lattice, and for the purpose of this discussion, $p$ is the loss rate).  The results  collapse  onto the universal scaling law 
\begin{equation}
\frac{\mu_{L}}{s_{p}}=\Phi\left(\frac{L^{2}}{s_p^{d/D}}\right)
\end{equation}
where $\Phi(x)=\chi \log( \xi \,x^{D/d}+1)$, shown as the solid line with fit parameters $\chi= 0.116$ and $\xi= 7.74$.

Intuitively, when the largest superplaquette is comparable to the size of the lattice, the effective diameter of the lattice is reduced, leading to an enhanced probability of attempting to match syndromes with  homologically non-trival paths.  This `interaction' between the loss rate and the computational error rate leads to non-universal scaling, and degrades the performance of the error correction.

\fig{largestSP}(right) shows the loss rate at which the largest superplaquette occupies half of the lattice, i.e. $p_{scale}$ is defined such that $\mu_{L}(p_{scale})=L^{2}/2$.  When $p_{loss}\geq p_{scale}$, the scaling law expected for small values of $p_{loss}$ breaks down.  When $L=16$, $p_{scale}=0.42$, which coincides with the point at which the red data points in \fig{PhaseDiagram} depart from simple quadratic behaviour.  When  $L=24$, $p_{scale}=0.44$, which coincides with the point at which the blue data points depart from the simple quadratic behaviour.

As $L\rightarrow\infty$, $p_{scale}\rightarrow1/2$, which is the percolation threshold.  Therefore, on larger lattices, the anomalous region, $p_{loss}>p_{scale}$ becomes smaller.



\section{Using degeneracy to improve the error correction threshold}
%

%

\begin{figure}
\begin{center}
\includegraphics[width=\columnwidth]{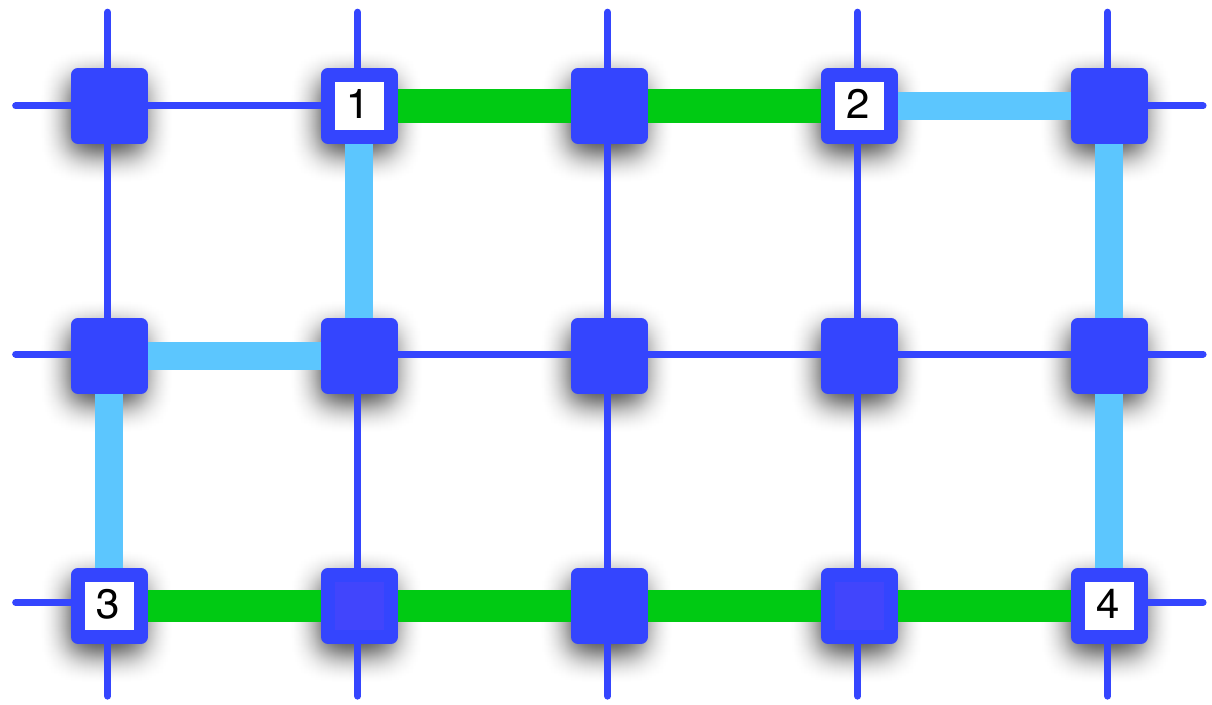}
\caption{There are two minimum weight matchings of the numbered nodes: $\{\{1,2\},\{3,4\}\}$ or $\{\{1,3\},\{2,4\}\}$.  The former has only one minimum weight matching path (green lines), whilst the latter has nine, (including the light blue line).} \label{parallelogram}
\end{center}
\end{figure}

The role of degeneracy in improving the performance of error correction algorithms has recently been noted by  Duclos-Cianci  and Poulin \cite{ducloscianci-2009}.  For a given configuration of the error syndromes, there may be a large number of degenerate minimum weight matchings.   \fig{parallelogram} exemplifies this for a configuration in which the eigenvalues of the four numbered plaquette have flipped.  There are two degenerate matchings of the numbered plaquettes for this case: $\{\{1,2\},\{3,4\}\}$ or $\{\{1,3\},\{2,4\}\}$.  The former matching has only a single \emph{matching path}, shown in green.  The latter matching has nine degenerate matching paths: there are three paths of minimum weight between 1 and 3, and likewise between 2 and 4.   The higher degeneracy of the latter matching indicates that it is more likely: there are more minimum-weight error chains that could have generated it.

When trying to find the most likely configuration of errors  given a certain error syndrome, this example demonstrates that one should account for the fact that some matchings have a higher degeneracy than others.  
 We now show that the path matching degeneracy can be handled by a small modification to algorithms based on Edmonds' perfect matching.

For a given matching, $M$, the degeneracy of the matching, $D_M$, is the number of shortest paths that are consistent with the matching.  In the case of a square lattice this degeneracy is simple to calculate, by considering the individual pairings within the matching $M=\{...,m_{ab},...\}$, where $m_{ab}$ is the pair of syndrome sites $\{a,b\}$.  For each pairing, $m_{ab}$ there are $h_{m_{ab}}$ horizontal edges and $v_{m_{ab}}$ vertical edges in any shortest path between the sites $a$ and $b$.  The set of all shortest paths between $a$ and $b$ is therefore the number of permutations of a string of $h_{m_{ab}}$ horizontal steps and $v_{m_{ab}}$ vertical steps
\begin{equation}
D_{m_{ab}}={h_{m_{ab}}+v_{m_{ab}} \choose v_{m_{ab}}} 
=\frac{(h_{m_{ab}}+v_{m_{ab}})!}{h_{m_{ab}}!v_{m_{ab}}!}.
\end{equation}
The matching degeneracy is then
\begin{equation}
D_M=\prod_{m\in M} D_m
\end{equation}

For the example shown in \fig{parallelogram}, the  matching $\{\{1,3\},\{2,4\}\}$ has pairings $\{1,3\}$ and $\{2,4\}$.  For the pair $\{1,3\}$ there is $h_{1,3}=1$ horizontal edge and $v_{1,3}=2$ vertical edges on any shortest path between the plaquettes. It follows that there are $D_{\{1,3\}}={3\choose 2}=3$ different shortest paths between 1 and 3. The same is true for the other pair  $\{2,4\}$:  $h_{2,4}=1$ and $v_{2,4}=2$, so $D_{\{2,4\}}=3$.  The degeneracy of the matching is then $D_{\{\{1,3\},\{2,4\}\}}=D_{\{1,3\}}D_{\{2,4\}}=9$.

We wish to find the matching that maximises the matching probability $P(M)$.  There are $D_M$ shortest matching paths, $\{E'_1,...,E'_{D_M}\}$, and since they are all of the same length, they each have a probability 
\begin{eqnarray}
P(E'_i)&=&\mathcal{N}\prod_{\forall\ell}e^{J_\ell u_\ell^{E'_i}},\nonumber\\
&=&\mathcal{N'} e^{-\sum_{\ell\in E'_i}2J_\ell},\nonumber\\
&=&\mathcal{N'} e^{-\sum_{m\in M}d_{m}},\nonumber\\
&=&P(E'_1),\nonumber
\end{eqnarray}
where $d_{m_{ab}}$ is the distance between sites $a$ and $b$.  In the absence of loss, $d_{m_{ab}}=2J({h_{m_{ab}}+v_{m_{ab}}})$.  The matching path probability is then
 \begin{eqnarray}
P(M)&=&\sum_{\{E'_i\}}P(E'_i),\nonumber\\
&=&D_M P(E'_1),\nonumber\\
&=&\mathcal{N'} e^{-\sum_{m\in M}(d_{m}-\ln D_m)}.\label{probM}
\end{eqnarray}
MaximIsing $P(M)$ is therefore equivalent to finding the matching that minimises 
\begin{equation}
G(M)=\sum_{m\in M}(d_{m}-\ln D_m).\label{G}
\end{equation}
This minimisation may be accomplished using the same methods described earlier, using Edmonds' perfect matching, where the distance between two sites $a$ and $b$ on the lattice is reduced by the amount $\ln D_{m_{ab}}$ to account for the number of degenerate paths between $a$ and $b$.

The form of \eqn{probM} suggests an analogy between the degeneracy and the entropic contribution to a  free energy, $G$, in a partition function.  A temperature scale does not appear explicitly in this expression: it is implicitly the `partition function' at $T=1$.\footnote{The analogy is not perfect, since we should include \emph{all}  matching paths  for $T>0$: whilst \eqn{probM} explicitly includes all matchings, it only accounts for the shortest matching paths consistent with a given matching.}  It follows that the matching that minimises \eqn{G} is not necessarily a shortest path. 

To make the analogy with a free energy more explicit, we modify \eqn{G} to
\begin{equation}
G_\tau(M)=\sum_{m\in M}(d_{m}-\tau \ln D_m),\label{tau}
\end{equation}
 interpreting $\tau$ as a temperature-like parameter.  When minimising $G_\tau$ in the limit $\tau\ll 2J\sim4$ (i.e.\ much less than the RBIM grain boundary tension) 
 excitations out of the degenerate ground state manifold are suppressed, so only microstates within the ground state manifold contribute to the entropy.  
 
\begin{figure}
\begin{center}
\includegraphics[width=\columnwidth]{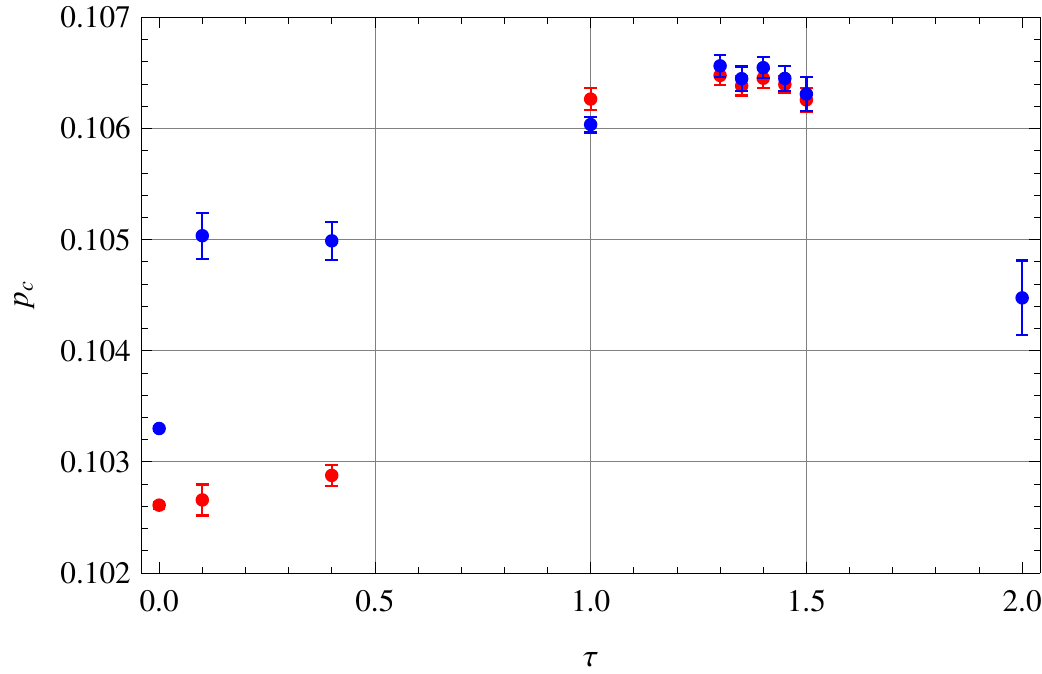}
\caption{The critical error probability as a function of the temperature parameter $\tau$.  The blue points were calculated on even  lattices with $L=16, 20, 24, 28$ and 32.  The red points were calculated on odd lattices with $L=15, 19, 23, 27$ and $31$.  The same error instances were used to calculate $p_c$ for all $0.1\leq\tau\geq1.5$, so the error bars are correlated.  The values at $\tau=0$ are taken from \cite{wang2003cht}.  The value at $\tau=2$ is shown to demonstrate the dramatic drop with increasing $\tau$.} \label{temp}
\end{center}
\end{figure}

 In any case, we can take $\tau$ to be a free parameter with which to tune the performance of the matching.  We can think of $\tau$ as parameterising a class of matching algorithms. In order to study the properties of this class of matching algorithms, 
 we compute the error correction threshold, $p_{c0}(\tau)$ for different values of $\tau$. \footnote{We note that all numerical results we report here in regard to the degeneracy are at $p_{loss}=0$.}  This is shown in \fig{temp}.\footnote{For each value of $\perr$, the same $10^6$ samples were used to estimate $p_{fail}$ for temperatures $0.1\leq\tau\leq1.5$.  As a result, the error bars in \fig{temp} are not uncorrelated.  Results at $\tau=0$ are taken from \cite{wang2003cht}.}

\begin{figure}
\begin{center}
\includegraphics[width=\columnwidth]{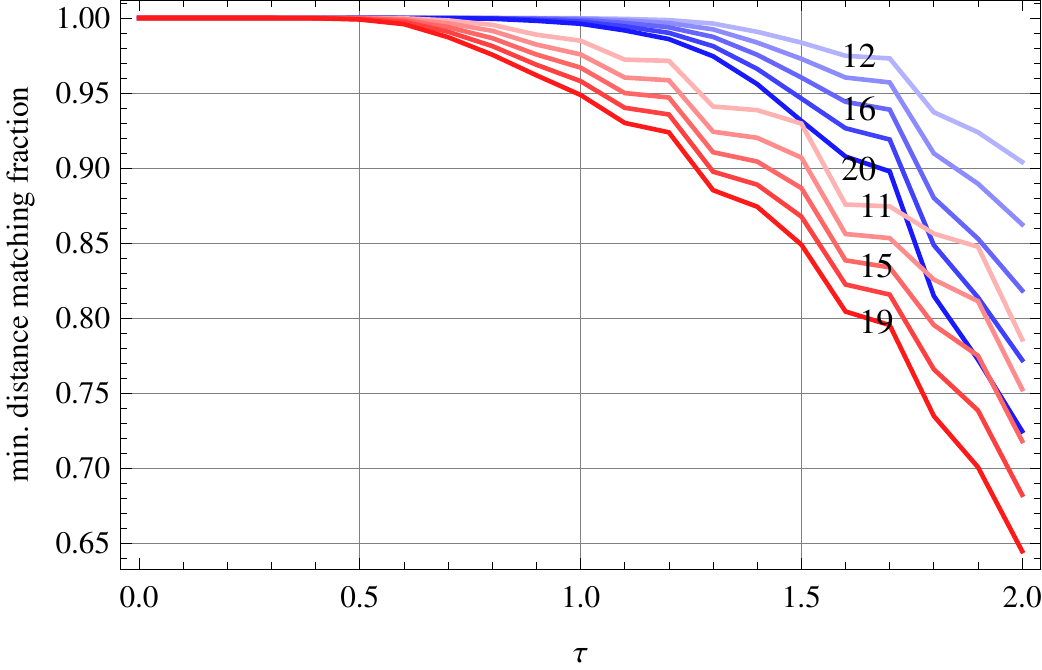}
\caption{The fraction of matchings that are minimum distance, for different values of $\tau$.  As $\tau$ increases, an increasing proportion of the selected matchings are longer than the minimum distance matchings (i.e. ground state matchings), but have sufficiently high degeneracy that their free energy is lower.  All simulations shown were done at $\perr=0.1035$, on different sized lattices, as indicated.  Note the disparity between the minimum distance matching fraction on even (blue) and odd (red) lattices.} \label{gsfrac}
\end{center}
\end{figure}

There are a number of conclusions to draw from  \fig{temp}.  Firstly, for $0<\tau\lesssim1.5$, the error threshold is higher than that at $\tau=0$.  For small $\tau$, the degeneracy in \eqn{tau} carries a small weight, so that $G$ is maximised with matchings in the ground state manifold: the matching algorithm chooses a maximally degenerate matching  \emph{within that manifold}.  This is demonstrated \fig{gsfrac}, where the fraction of matchings that are also minimum distance matchings is shown for different values of $\tau$.  For $\tau<0.5$, this fraction is essentially unity, so that all matchings are also minimum distance matchings.

The threshold increases with increasing $\tau$, and peaks around $\tau=1.4$ or so.  Interestingly, this is somewhat better than the threshold at $\tau=1$, pertaining to \eqn{G}.  When $\tau\gtrsim1$, the matching algorithm no longer simply maximises the matching path probability, $P(M)$.  Instead, it favours more degenerate paths.   Finally, the threshold decreases as $\tau$ increases further.     This behaviour is qualitatively consistent with the proposed re-entrant phase diagram for the RBIM, shown in e.g.\ \cite{wang2003cht}.

There is a notable discrepancy between the thresholds and minimum distance matching fractions on even and odd lattices.  For $\tau<1$ even lattices appear to have a substantially higher threshold than do odd lattices. This discrepancy is much larger than that discussed in \cite{wang2003cht}, shown for $\tau=0$.  The deviation of the minimum weight matching fraction from unity also follows different trends depending on the parity of the lattice: on odd lattices it begins deviating from unity for $\tau\gtrsim0.6$, whilst  on even lattices the deviation sets in at $\tau\gtrsim1$ (see \fig{gsfrac}).  Because the system is discrete, there is necessarily a gap between the ground state manifold and the low lying excited states of the RBIM.  In particular, \fig{gsfrac}, suggests that this gap is smaller on odd lattices than on even lattices, so that at a given `temperature' $\tau$ there a higher chance of that an odd lattice will be in an excited state.  The origin of this discrepancy remains an open problem.

Given the maximum possible threshold is set by the Nishimorit point at $\sim10.9\%$ \cite{wang2003cht}, the addition of degeneracy in the matching algorithm improves the threshold by a modest but statistically significant margin.


\section{What can we  really say about the \emph{RBIM} phase diagram?}
%

%


The most precise estimate for $p_{c0}$, the $T=0$ phase transition in the RBIM, is currently the value reported in \cite{wang2003cht}, and this result has been applied in the study of the classical 2D RBIM.  The observation of the huge degeneracy in the ground state manifold of the RBIM raises the question of how to ensure that the ground states (i.e.\ minimum-distance matchings) selected by Edmonds' algorithm are sampled \emph{fairly} from the manifold.

  As an elementary test, we applied two freely available implementations,  \emph{Blossom IV} \cite{WilliamCook01011999}  and \emph{Blossom V} \cite{kolmogorov2009blossom},  to the simple matching problem depicted in \fig{parallelogram}, 
permuting vertex labels through all 24 permutations to reorder the presentation of the complete graph.  Regardless of presentation, \emph{Blossom IV} always returned the least degenerate matching, $\{\{1,2\},\{3,4\}\}$, regardless of reordering.  On the other hand, depending on the vertex labels, \emph{Blossom V}  returned both possible matchings: half of the  permutations resulted in matching $\{\{1,2\},\{3,4\}\}$, and the other half yielded the alternative,  $\{\{1,3\},\{2,4\}\}$.  This test serves to demonstrate that different implementations of Edmonds' matching sample amongst the ground states differently.

To compute $p_{c0}$ accurately, we need to sample the ground states of the RBIM fairly, i.e.\ with a probability distribution proportional to the path matching degeneracy.  To see this, we define an observable $\mathcal O_{E,E'}$ for the classical RBIM to be the indicator function determining whether $E+E'$ is in the trivial homology class:
\begin{equation}
\mathcal O_{E,E'}=\left\{ \begin{array}{cl}0 & \quad\textrm{if $E+E'$ is homologically trivial}  \\1 & \quad\textrm{otherwise}\end{array}\right.\nonumber
\end{equation}
 As in \cite{wang2003cht}, $p_{c0}$ is calculated from the scaling behaviour of  $p_{fail}=\langle\mathcal O_{E,E'}\rangle$, which is given by:
\begin{eqnarray}
\langle\mathcal O_{E,E'}\rangle&=&\sum_EP(E)\langle\mathcal O_{E,E'}\rangle_E\nonumber\\
&=&\lim_{\beta\rightarrow0}\sum_EP(E)\sum_{\forall\sigma}\mathcal O_{E,\sigma}e^{-\beta H_{E,\sigma}} /Z\nonumber\\
&=&\sum_EP(E)\sum_{\sigma\in \mathrm{g.s.}}\mathcal O_{E,\sigma}/\mathcal D_E\, \nonumber\\
&=&\sum_EP(E)\sum_{E'\in \mathrm{g.s.}}\mathcal O_{E,E'}/\mathcal D_E\,\label{fairsampling} \\
&=&\sum_EP(E)\sum_{M'}\mathcal O_{E,M'} D_{M'}/\mathcal D_E\label{matchingdegen}
\end{eqnarray}
where $D_{M'}$ is the matching path degeneracy of the matching $M'$, $\mathcal D_E=\sum_{M'|\partial E} D_M'$ is the ground state (g.s.) degeneracy of the RBIM instance $E$, the grain boundary $E'$ is interchangeable with a given Ising spin configuration $\sigma$ (as described above, and in \cite{dennis:4452}), and 
\begin{equation}
H_{E,\sigma}=\sum_{\langle   i j\rangle}{\beta J_{ij} u_{ij}^{E}\sigma_i\sigma_j}.
\end{equation}
Also we have used the fact that $O_{E,E'}=O_{E,M'}$ is constant over all minimum weight matching paths $E'$ consistent with a given matching $M'$.   Note that in the limit $L\rightarrow\infty$, the sum over $E$ is made unnecessary by appealing to an ergodic argument:  $\langle\mathcal O\rangle= \langle\mathcal O\rangle_E$, for any $E$ whose errors are uniformly distributed.

The final result, \eqn{matchingdegen}, establishes the distribution from which we should sample the matchings: $P(M|E)=D_M/\mathcal{D}_E$.  This result suggests that the direct use of existing implementations of Edmonds' perfect matching may not give accurate estimates of $p_{c0}$.

  In principle, for a finite lattice, one could enumerate \emph{all} matchings, compute the matching-path degeneracy for each, then evaluate \eqn{matchingdegen} explicitly, however this is computationally inefficient.  We do not know of a computationally efficient way to ensure a fair sampling.
  
 We note however, that within the ground state manifold (i.e.\ taking $0<\tau<0.5$), the error correction threshold is bounded by about 0.105 (see \fig{temp}).  This suggests that $p_{c0}\leq0.105$.  That this is statistically less than the Nishimori probability $p_c$ gives support to the notion that the phase diagram is indeed reentrant, as has been discussed elsewhere.  However we propose that an accurate value of $p_{c0}$ may be higher than that reported in \cite{wang2003cht}.

\section{conclusion}


We have shown using both numerical studies and analytic results that the simultaneous expression of both loss and computational errors can be dealt with in the context of topological memories.  The trade-off between these processes is rather gentle.  At low loss rates, our analytical results indicate that, because superplaquettes embody more physical qubits, the principle effect of losses is a moderate, linear increase in the effective error rate per plaquette.  At higher loss rates, changes in the structure and connectivity of the superplaquete lattice become significant.

We have also shown that the classical algorithms used to implement topological quantum error correction can be improved by accounting for degeneracy in matching algorithms.  This can take various forms; either in modifying edge weights between neighbouring superplaquettes on an irregular lattice as discussed in \cite{stace:200501}, or by counting the number of degenerate shortest-paths between syndromes, as discussed here and in \cite{ducloscianci-2009}.  Both approaches may be implemented with little overhead to existing matching routines.

This work also highlights the attention that needs to be given to degeneracy when calculating thresholds in the RBIM and any related classical statistical mechanics simulations relying on implementations of Edmonds' matching algorithm.  To accurately predict $p_{c0}$, minimum-distance matchings should be sampled \emph{fairly} from amongst the hugely degenerate ground state manifold.  To our knowledge, achieving this remains an open  problem.  We hope that this avenue of research will shed new light on the RBIM.

TMS thanks the Australian Research Council for funding.  SDB thanks the EPSRC and the CQCT for funding.  We thank  Ben Powell, Andrew Doherty, Bill Cook and Vladimir Kolmogorov for useful discussions. We also thank Jim Harrington for extensive conversations, and for conducting the numerical simulations whose results are  incorporated in Figures \ref{temp} and \ref{gsfrac}.


\end{document}